\documentclass[pra,aps,showpacs,twocolumn]{revtex4}

\usepackage{amsmath}
\usepackage{bm}
\usepackage{graphicx}

\begin{document}

\title{Magnetization of a half-quantum vortex in a spinor Bose-Einstein
condensate}

\author{Shoichi Hoshi and Hiroki Saito}

\affiliation{Department of Applied Physics and Chemistry,
The University of Electro-Communications, Tokyo 182-8585, Japan}

\date{\today}

\begin{abstract}
Magnetization dynamics of a half-quantum vortex in a spin-1 Bose-Einstein
condensate with a ferromagnetic interaction are investigated by mean-field
and Bogoliubov analyses.
The transverse magnetization is shown to break the axisymmetry and form
threefold domains.
This phenomenon originates from the topological structure of the
half-quantum vortex and spin conservation.
\end{abstract}


\maketitle

\section{Introduction}

The spin degrees of freedom in an atomic Bose-Einstein condensate (BEC)
allow for a variety of topological defects.
Mermin-Ho~\cite{Mermin} and Anderson-Toulouse~\cite{Anderson} vortices
have been produced by the MIT group~\cite{Leanhardt03} using the
topological phase-imprinting method~\cite{Leanhardt02}.
Recently, the Berkeley group observed spontaneous formation of polar-core
vortices in magnetization of a spin-1 $^{87}{\rm Rb}$ BEC~\cite{Sadler}.
In addition, several theoretical predictions have been made of various
topological defects in spinor BECs, such as fractional
vortices~\cite{Volovik,Leonhardt,Zhou},
skyrmions~\cite{Khawaja}, and knot structures~\cite{Kawaguchi}.

Topological-defect structures in spinor BECs are closely related to the
symmetry groups of spin states~\cite{Ho}.
For example, the symmetry group of the polar state ($m = 0$, where $m$
denotes the magnetic sublevel of the spin) in a spin-1 BEC is ${\rm U(1)}
\times S^2 / {\rm Z}_2$, and distinct configurations of the polar state
can be specified by elements of this group.
This symmetry group allows the topological structure of the half-quantum
vortex~\cite{Volovik,Leonhardt,Zhou}.
On the other hand, the symmetry group of the ferromagnetic state ($|m| =
1$) in a spin-1 BEC is SO(3).

Let us consider a half-quantum vortex prepared in an antiferromagnetic
BEC.
We then consider a magnetic phase transition occurring from the
antiferromagnetic to ferromagnetic phases.
Since the symmetry group of the spin state changes from ${\rm U(1)} \times
S^2 / {\rm Z}_2$ to SO(3) and the half-quantum vortex structure cannot
exist for the latter symmetry group, the ensuing magnetization dynamics
are expected to break the symmetry and create nontrivial states.
Such a change of spin texture associated with a change of the spin state
symmetry group is the subject of the present paper.

In the present paper, we study the magnetization dynamics of the
half-quantum vortex state produced in a spin-1 ferromagnetic BEC.
The initial state can be prepared by, e.g., the phase-imprinting method
using Laguerre-Gaussian beams~\cite{Andersen}.
We show that the half-quantum vortex develops into threefold magnetic
domains through dynamical instability.
This contrasts with the magnetization of a uniform polar state, in which
a polar-core vortex or twofold domain structure is formed~\cite{SaitoL}.
We study the dynamical instability using the Bogoliubov analysis,
numerically in a 2D system and analytically in a 1D ring.
The threefold domain formation can also be understood from the topological
structure of a half-quantum vortex and spin conservation, and its
geometrical interpretation is provided.

This paper is organized as follows.
Section~\ref{s:formulation} introduces the formulation of the problem and
defines the notation used.
Section~\ref{s:dynamics} shows the magnetization dynamics of the
half-quantum vortex and demonstrates threefold domain formation.
Section~\ref{s:Bogo} details the Bogoliubov analysis to study the
dynamical instability.
Section~\ref{s:geometry} is devoted to the geometrical interpretation of
the threefold domain formation.
Section~\ref{s:conclusion} gives the conclusions to the study.

\section{Formulation of the problem}
\label{s:formulation}

We consider a BEC of spin-1 atoms with mass $M$ confined in an
axisymmetric harmonic potential $V(\bm{r}) = M [\omega_\perp^2 (x^2 + y^2)
+ \omega_z^2 z^2] / 2$ that is independent of the magnetic sublevels $m$
of the spin.
In the mean-field approximation, the condensate can be described by the
macroscopic wave functions $\psi_m(\bm{r})$ with $m = -1, 0, 1$ satisfying
\begin{equation} \label{normal}
\int d\bm{r} \sum_{m = -1}^1 |\psi_m|^2 \equiv \int d\bm{r} \rho = N,
\end{equation}
where $N$ is the number of atoms.
The magnetization density is given by
\begin{equation} \label{Fdef}
\bm{F} = \sum_{m, m'} \psi_m^* \bm{f}_{m m'} \psi_{m'},
\end{equation}
where $\bm{f}$ is the vector of the spin-1 matrices.
The macroscopic wave functions at zero temperature obey the
three-component Gross-Pitaevskii (GP) equations,
\begin{subequations} \label{GP}
\begin{eqnarray}
i\hbar \frac{\partial \psi_0}{\partial t} & = &
\left( -\frac{\hbar^2}{2M} \nabla^2 + V + c_0 \rho \right) \psi_0
\nonumber \\
& & + \frac{c_1}{\sqrt{2}} \left( F_+ \psi_1 +  F_- \psi_{-1} \right),
\\
i\hbar \frac{\partial{\psi_{\pm 1}}}{\partial t} & = & \left(
-\frac{\hbar^2}{2M}{\nabla}^2 + V + c_0\rho \right) \psi_{\pm 1}
\nonumber \\
& & + c_1 \left( \frac{1}{\sqrt{2}} F_\mp \psi_0 \pm F_z \psi_{\pm 1}
\right),
\end{eqnarray}
\end{subequations}
where $F_\pm = F_x \pm i F_y$.
The interaction coefficients in Eq.~(\ref{GP}) are defined as
\begin{equation}
c_0 = \frac{4 \pi \hbar^2}{M} \frac{a_0 + 2 a_2}{3},
\qquad
c_1 = \frac{4 \pi \hbar^2}{M} \frac{a_2 - a_0}{3},
\end{equation}
with $a_0$ and $a_2$ being the $s$-wave scattering lengths for colliding
channels with total spins 0 and 2.
For $c_1 < 0$, the ferromagnetic state is energetically favorable, while
for $c_1 > 0$ the polar or antiferromagnetic state is favorable.
In the present paper, we restrict ourselves to spin-1 $^{87}{\rm Rb}$
atoms, which have a positive $c_0$ and a negative $c_1$~\cite{Schmal}.

We assume that $\hbar \omega_z$ is much larger than other characteristic
energies and the condensate has a tight pancake shape.
The condensate wave function is therefore frozen in the ground state of
the harmonic potential in the $z$ direction and the system is effectively
2D.
Integrating the GP energy functional with respect to $z$, we find that the
2D wave function $\psi_m^{\rm 2D}$ follows the GP equation having the same
form as Eq.~(\ref{GP}), where the interaction coefficients $c_0$ and $c_1$
are multiplied by $[m \omega_z / (2 \pi \hbar)]^{1/2}$.
We define a normalized wave function,
\begin{equation}
\tilde\psi_m = \frac{1}{\sqrt{N}} \frac{\hbar}{m \omega_\perp} \psi_m^{\rm
2D},
\end{equation}
and normalized interaction coefficients,
\begin{equation}
\tilde c_j = \frac{N}{\hbar \omega_\perp} \sqrt{\frac{m \omega_z}{2 \pi
\hbar}} \frac{m \omega_\perp}{\hbar} c_j
\end{equation}
with $j = 0$ and $1$.
For example, using the scattering lengths of a spin-1 $^{87}{\rm Rb}$
atom $a_0 = 101.8 a_{\rm B}$ and $a_2 = 100.4 a_{\rm B}$~\cite{Kempen},
where $a_{\rm B}$ is the Bohr radius, and trap frequencies $\omega_\perp =
2\pi \times 200$ Hz and $\omega_z = 2\pi \times 4$ kHz, the interaction
coefficients become
\begin{equation}
\tilde c_0 \simeq 0.16 N, \;\;\;\;
\tilde c_1 \simeq -\tilde c_0 / 216.
\end{equation}

We consider a half-quantum vortex state given by~\cite{Leonhardt}
\begin{equation} \label{hqv}
\left( \begin{array}{c}
						\psi_1^{\rm hqv} \\
						\psi_0^{\rm hqv} \\
						\psi_{-1}^{\rm hqv}
						\end{array}
				\right) = \left( \begin{array}{c}
						{f_1(r)}e^{\pm i \theta} \\
						0 \\
						{f_{-1}(r)}
						\end{array}
				\right),
\end{equation}
where $r = (x^2 + y^2)^{1/2}$ and $\theta = {\rm arg}(x + i y)$.
The functions $f_{\pm 1}(r)$ are stationary solutions of Eq.~(\ref{GP})
satisfying
\begin{equation} \label{half}
\int_0^\infty 2 \pi r^2 |f_{\pm 1}(r)|^2 dr = \frac{N}{2}.
\end{equation}
From this condition, the state (\ref{hqv}) has an angular momentum of $N
\hbar / 2$.
Without loss of generality, we restrict ourselves to the upper sign in
Eq.~(\ref{hqv}) unless otherwise stated.

As several authors have discussed~\cite{Volovik,Leonhardt,Zhou}, the
half-quantum vortex has an interesting topological structure.
Equation (\ref{hqv}) is invariant under the transformation $\exp(i \phi
/ 2) \exp(i f_z \phi / 2) \exp(-\phi \partial_\theta)$, where $\phi$
is an arbitrary angle and $f_z = m$ for $\psi_m$.
This indicates that spatial rotation by an angle $\phi$ around the $z$
axis is accompanied by spin rotation by $-\phi / 2$ with an additional
phase factor $\exp(i \phi / 2)$.
Thus, for a rotation around the $z$ axis by $2\pi$, the spin rotates
only by $-\pi$.

Thermodynamic stability of the half-quantum vortex is studied in
Ref.~\cite{Isoshima}, while a half-quantum vortex ring is discussed in
Ref.~\cite{Ruo}.
Recently, it has been predicted that half-quantum vortices can be
nucleated in rotating traps~\cite{Ji,Chiba} and fluctuation-drive vortex
fractionalization has been proposed in Ref.~\cite{Song}.

\section{Magnetization dynamics of a half-quantum vortex}
\label{s:dynamics}

In this section, we study the magnetization dynamics of the half-quantum
vortex state (\ref{hqv}) for a spin-1 $^{87}{\rm Rb}$ BEC.

The initial state is assumed to be the half-quantum vortex state
(\ref{hqv}) obtained by the imaginary-time propagation method.
An experimental method to realize this initial state is discussed later.
It follows from Eq.~(\ref{GP}) that when $\psi_0$ is exactly zero as in
Eq.~(\ref{hqv}), $\psi_0$ always vanishes in the mean-field evolution and
no magnetization occurs even for a ferromagnetic interaction.
We therefore add a small amount of initial noise to $\psi_0$, which
triggers growth of the $m = 0$ component. 
Physically, this initial noise corresponds to quantum and thermal
fluctuations and experimental imperfections~\cite{SaitoL,SaitoA,SaitoKZ}.
We set the noise as $\tilde \psi_0 = r_1 + i r_2$ on each mesh point,
where $r_1$ and $r_2$ are random numbers with uniform distribution between
$\pm 10^{-3}$.
For the imaginary- and real-time propagations, we employ the
Crank-Nicolson method with the size of each mesh being 0.05 $\sqrt{\hbar /
(m \omega_\perp)}$.

\begin{figure}[t]
\includegraphics[width=8cm]{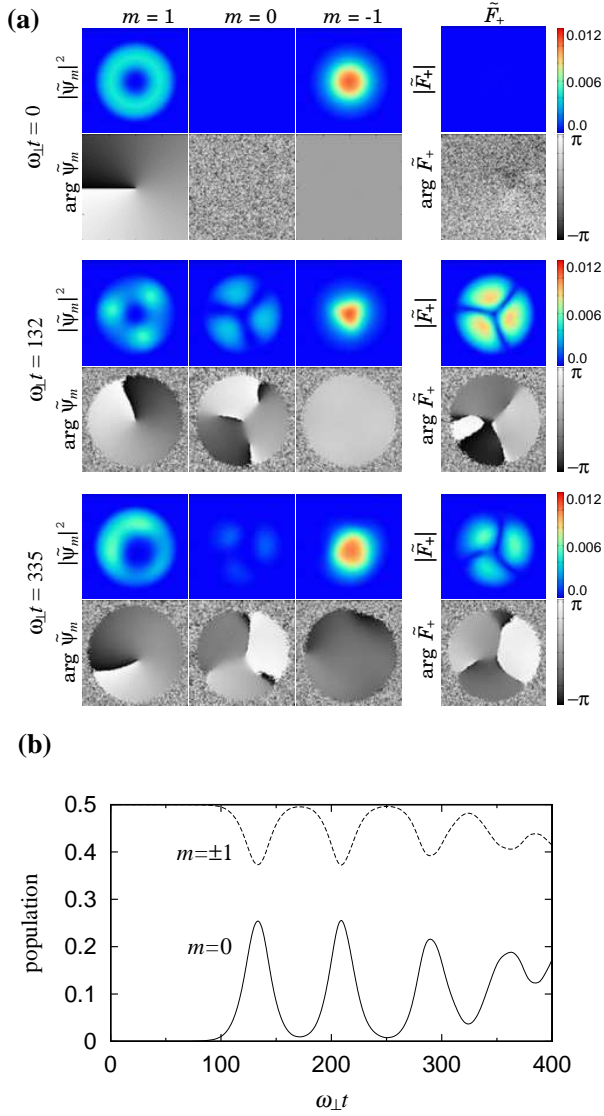}
\caption{
(color) (a) Time evolution of the normalized wave function $\tilde\psi_m$
and transverse magnetization $\tilde F_+ = \tilde F_x + i \tilde F_y$,
where $\tilde F_x$ and $\tilde F_y$ are defined by Eq.~(\ref{Fdef}) with
$\psi_m$ replaced by $\tilde \psi_m$.
The size of each panel is $10 \times 10$ in units of $\sqrt{\hbar / (m
\omega_\perp)}$.
(b) Time evolution of the population in each component.
The interaction coefficients are $\tilde c_0 = 3000$ and $\tilde c_1 =
-\tilde c_0 / 216$ (corresponding to spin-1 $^{87}{\rm Rb}$).
}
\label{f:dynamics}
\end{figure}
Figure \ref{f:dynamics} (a) shows the time evolution of the density and
phase profiles of each spin component and that of the transverse
magnetization.
At $\omega_\perp t = 0$, the transverse magnetization almost vanishes
because of the initial state (\ref{hqv}) with small noise added in the $m
= 0$ component.
The $m = 0$ component then grows in time and exhibits a threefold pattern,
which leads to threefold magnetic domains as shown in
Fig.~\ref{f:dynamics} (a) ($\omega_\perp t = 132$).
This threefold domain formation is the main result of the present paper.
We note that the magnetization in the three domains have different
directions to cancel and conserve the total spin.
Then, the population for each $m$ oscillates as shown in
Fig.~\ref{f:dynamics} (b) due to the excess energy, and the appearance and
disappearance of the threefold domains are repeated.
At $\omega_\perp t \simeq 300$, the instability in the dipole mode becomes
significant and the system undergoes a dipole deformation as shown in
Fig.~\ref{f:dynamics} (a), which is followed by complicated dynamics.
Throughout the dynamics, the total density profile is almost unchanged and
remains in the Thomas-Fermi distribution.
The threefold domains are generated for both signs in the initial state
(\ref{hqv}).
We checked that no dynamics occur for an antiferromagnetic interaction
($c_1 > 0$).

We now discuss how to prepare the initial state in an experiment.
First we create a non-rotating BEC in the $m = -1$ component,
$\psi_{-1}^{\rm ini}$, and then Gaussian and Laguerre-Gaussian beams
propagating in the same direction are applied.
The frequencies of these beams are tuned to the Raman transition from the
non-rotating $m = -1$ state to the $m = 1$ state with a unit angular
momentum~\cite{Andersen}.
We thus obtain the half-quantum vortex state given by
\begin{equation}
\left( \begin{array}{c} \psi_1 \\ \psi_0 \\ \psi_{-1} \end{array} \right)
= \left( \begin{array}{c} \psi_{-1}^{\rm ini} e^{i \theta} \sin (A r e^{-B
r^2}) \\ 0 \\ \psi_{-1}^{\rm ini} \cos (A r e^{-B r^2}) \end{array}
\right), 
\end{equation}
where $A$ and $B$ are proportional to the intensity and width of the
beams, respectively.
For example, $A = 0.196 (m \omega_\perp / \hbar)^{1/2}$ and $B = 0.002 m
\omega_\perp / \hbar$ give density profiles similar to the initial state
in Fig.~\ref{f:dynamics}.
We have confirmed that the dynamics from this initial state are
qualitatively the same as those in Fig.~\ref{f:dynamics}.
The pattern formation as shown in Fig.~\ref{f:dynamics} (a) can be
observed using the Stern-Gerlach separation and a nondestructive
spin-sensitive measurement~\cite{Higbie}.

\section{Bogoliubov analysis}
\label{s:Bogo}

\subsection{Numerical diagonalization}
\label{s:numbogo}

The dynamics shown in Fig.~\ref{f:dynamics} suggest that the half-quantum
vortex state has dynamical instabilities.
In this section, we perform the Bogoliubov analysis.

We decompose the macroscopic wave functions $\psi_m$ into the stationary
state $\psi_m^{\rm hqv}$ in Eq.~(\ref{hqv}) and small deviations $\delta
\psi_m$ from this state as
\begin{equation} \label{psiB}
\left( \begin{array}{c} \psi_1 \\ \psi_0 \\ \psi_{-1} \end{array} \right)
= \left( \begin{array}{c} e^{-i \mu_1 t / \hbar} (\psi_1^{\rm hqv} +
\delta \psi_1) \\ e^{-i (\mu_1 + \mu_{-1}) t / (2\hbar)} \delta \psi_0 \\
e^{-i \mu_{-1} t / \hbar} (\psi_{-1}^{\rm hqv} + \delta \psi_{-1}) 
\end{array} \right),
\end{equation}
where the chemical potential in each component is defined by
\begin{equation} \label{mu}
\mu_{\pm 1} = \frac{2}{N} \int d\bm{r} \left[ \psi_{\pm 1}^{{\rm hqv} *}
\left( -\frac{\hbar^2}{2M} \nabla^2 + V + c_0 \rho \pm c_1 F_z \right)
\psi_{\pm 1}^{\rm hqv} \right].
\end{equation}
Here $\rho$ and $F_z$ are given by Eqs.~(\ref{normal}) and (\ref{Fdef})
with $\psi_m$ replaced by $\psi_m^{\rm hqv}$.
Substituting Eq.~(\ref{psiB}) into Eq.~(\ref{GP}), we obtain the
Bogoliubov-de Gennes equations:
\begin{subequations}
\label{Bogo}
\begin{eqnarray}
i \hbar \frac{\partial \delta \psi_0}{\partial t} & = &
\left( -\frac{\hbar^2}{2M} \nabla^2 + V - \frac{\mu_1 + \mu_{-1}}{2}
\right) \delta \psi_0
\nonumber \\
& & + (c_0 + c_1) \left( |\psi_1^{\rm hqv}|^2 +
|\psi_{-1}^{\rm hqv}|^2 \right) \delta \psi_0 \nonumber \\
& & + 2 c_1 \psi_1^{\rm hqv} \psi_{-1}^{\rm hqv} \delta \psi_0^*,
\label{B1} \\
i \hbar \frac{\partial \delta \psi_{\pm 1}}{\partial t} & = &
\left( -\frac{\hbar^2}{2M} \nabla^2 + V - \mu_{\pm 1} \right) \delta
\psi_{\pm 1}
\nonumber \\
& & + (c_0 + c_1) \left[ 2 |\psi_{\pm 1}^{\rm hqv}|^2 \delta
\psi_{\pm 1} + (\psi_{\pm 1}^{\rm hqv})^2 \delta \psi_{\pm 1}^* \right]
\nonumber \\
& & + (c_0 - c_1) \bigl( |\psi_{\mp 1}^{\rm hqv}|^2 \delta \psi_{\pm 1}
+ \psi_{\mp 1}^{{\rm hqv} *} \psi_{\pm 1}^{\rm hqv} \delta \psi_{\mp 1}
\nonumber \\
& & + \psi_1^{\rm hqv} \psi_{-1}^{\rm hqv} \delta \psi_{\mp 1}^* \bigr),
\label{B2}
\end{eqnarray}
\end{subequations}
where we take only the first order of $\delta \psi_m$.
We note that both Eqs.~(\ref{B1}) and (\ref{B2}) have a closed form within
an angular-momentum subspace.
Expanding $\delta \psi_m$ as
\begin{equation} \label{expand}
\delta \psi_m = \sum_\ell \left[ \alpha_\ell^{(m)}(r, t) +
\beta_\ell^{(m)}(r, t) \right] e^{i \ell \theta},
\end{equation}
we find that $\alpha^{(0)}_\ell$ only couples with
$\beta^{(0)}_{1-\ell}$, and $\alpha^{(1)}_\ell$ only couples with
$\beta^{(1)}_{2-\ell}$, $\alpha^{(-1)}_{\ell-1}$, and
$\beta^{(-1)}_{1-\ell}$.
We therefore define the modes as
\begin{equation} \label{mode0}
\delta \psi_{0,\ell} = \alpha^{(0)}_\ell(r) e^{i \ell \theta} e^{-i \omega
t} + \beta^{(0)}_{1-\ell}(r) e^{i (1 - \ell) \theta} e^{i \omega t},
\end{equation}
for the $m = 0$ component and
\begin{subequations} \label{mode1}
\begin{eqnarray}
\delta \psi_{1, \ell} & = & \alpha_\ell^{(1)}(r) e^{i \ell \theta} e^{-i
\omega t} + \beta_{2-\ell}^{(1)}(r) e^{i (2 - \ell) \theta} e^{i \omega
t}, \\
\delta \psi_{-1, \ell} & = & \alpha_{\ell-1}^{(-1)}(r) e^{i (\ell - 1)
\theta} e^{-i \omega t} + \beta_{1-\ell}^{(-1)}(r) e^{i (1 - \ell) \theta}
e^{i \omega t},
\end{eqnarray}
\end{subequations}
for the $m = \pm 1$ components.

\begin{figure}[t]
\includegraphics[width=9cm]{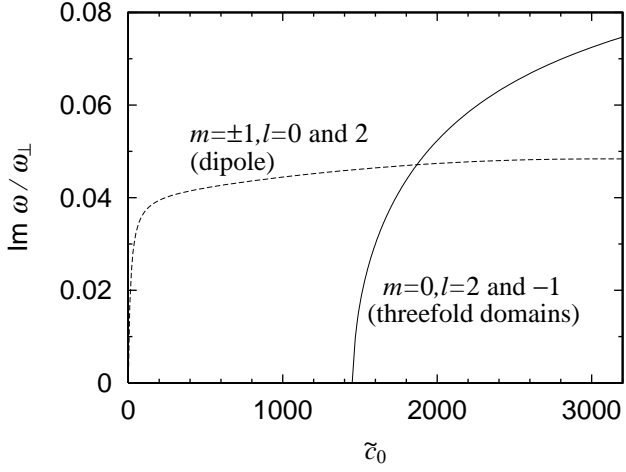}
\caption{
Imaginary part of the Bogoliubov spectrum for the half-quantum vortex
state of a spin-1 $^{87}{\rm Rb}$ BEC ($\tilde c_1 = -\tilde c_0 / 216$).
The solid line is obtained by diagonalizing Eq.~(\ref{B1}), where the mode
function has the form (\ref{mode0}) with  $\ell = 2$ and $-1$,
corresponding to the threefold domain formation.
The dashed line is obtained from Eq.~(\ref{B2}), where the mode functions
have the form (\ref{mode1}) with $\ell = 0$ and $2$, corresponding to the
dipole deformation.
}
\label{f:Bogo}
\end{figure}
We numerically diagonalize Eq.~(\ref{Bogo}) using the method in
Ref.~\cite{Edwards}.
Figure~\ref{f:Bogo} shows the imaginary part of the Bogoliubov spectrum
for various values of $\tilde c_0$ and $\tilde c_1$ with $\tilde c_1 =
-\tilde c_0 / 216$.
Diagonalizing Eq.~(\ref{B1}) for the $m = 0$ component, we find that the
excitation energies of the modes (\ref{mode0}) with $\ell = 2$ and $\ell =
-1$ have an imaginary part for $\tilde c_0 \gtrsim 1450$.
Diagonalization of Eq.~(\ref{B2}) shows that the modes (\ref{mode1}) with
$\ell = 0$ and $\ell = 2$ are dynamically unstable for $\tilde c_0 > 0$.
The imaginary part for the $m = 0$ modes exceeds that for the $m = \pm 1$
modes at $\tilde c_0 \simeq 1870$, and the growth in the $m = 0$ component
becomes dominant for $\tilde c_0 \gtrsim 1870$.

When the $m = 0$ modes with $\ell = 2$ and $\ell = -1$ grow due to the
dynamical instability, $|\delta \psi_0|^2$ becomes
\begin{eqnarray} \label{three}
\left|\delta \psi_0 \right|^2 & \simeq & \left| \delta \psi_{0, \ell = 2}
+ \delta \psi_{0, \ell = -1} \right|^2 \nonumber \\
& = & \left| \alpha^{(0)}_2 + \beta^{(0)}_2 \right|^2 + \left|
\alpha^{(0)}_{-1} + \beta^{(0)}_{-1} \right|^2 
\nonumber \\
& & + 2 \left|
\left(\alpha^{(0)}_2 + \beta^{(0)}_2\right) \left(\alpha^{(0)}_{-1} +
\beta^{(0)}_{-1}\right) \right| \cos (3\theta + \delta), \nonumber \\
\end{eqnarray}
where $\delta = {\rm arg} [ (\alpha^{(0)}_2 + \beta^{(0)}_2)
(\alpha^{(0)}_{-1} + \beta^{(0)}_{-1})^* ]$. 
We can show that $|F_+|^2$ also has a similar form as a function of
$\theta$.
Equation (\ref{three}) indicates that these dynamically unstable modes
generate threefold domains, in agreement with the result in
Fig.~\ref{f:dynamics} (a).
The dynamically unstable modes of the $m = \pm 1$ components have angular
momenta $1 \pm 1$ in the $m = 1$ component and $0 \pm 1$ in the $m = -1$
component.
These modes therefore correspond to dipole deformation, which again
explains the result in Fig.~\ref{f:dynamics} (a).
Since the imaginary part of the $m = 0$ excitation energy is larger than
that of $m = \pm 1$ for $\tilde c_0 = 3000$, the threefold domains first
emerge, followed by the dipole deformation.

\subsection{1D ring model}
\label{s:ring}

For simplicity, we analyze a 1D ring model in order to understand the
dynamical instabilities in Fig.~\ref{f:Bogo}.

We assume that the system is confined in a 1D ring with radius $R$, and
the effective interaction coefficients are denoted by $c_0^{\rm 1D}$ and
$c_1^{\rm 1D}$.
For the stationary state $\psi_m^{\rm hqv}$ in Eq.~(\ref{psiB}), we take
\begin{equation} \label{1Dhqv}
\psi_1^{\rm hqv} = \sqrt{\frac{n}{2}} e^{i \theta}, \qquad
\psi_{-1}^{\rm hqv} = \sqrt{\frac{n}{2}},
\end{equation}
where $n = N / (2 \pi R)$ is the atomic density and $\theta$ is the
azimuthal angle.
The chemical potentials in Eq.~(\ref{mu}) read
\begin{equation} \label{1Dmu}
\mu_1 = K + c_0^{\rm 1D} n, \qquad
\mu_{-1} = c_0^{\rm 1D} n,
\end{equation}
where
\begin{equation}
K \equiv \frac{\hbar^2}{2 M R^2}.
\end{equation}
Substituting Eqs.~(\ref{1Dhqv}) and (\ref{1Dmu}) into Eq.~(\ref{B1}) gives
\begin{equation}
i \hbar \frac{\partial \delta \psi_0}{\partial t} =
\left( -K \frac{d^2}{d \theta^2} - \frac{K}{2} + c_1^{\rm 1D} n \right)
\delta \psi_0 + c_1^{\rm 1D} n e^{i \theta} \delta \psi_0^*.
\end{equation}
Assuming that $\delta \psi_0$ has the form
\begin{equation}
\delta \psi_0 = \alpha_\ell^{(0)} e^{i \ell \theta} e^{-i \omega t} +
\beta_{1-\ell}^{(0)} e^{i (1 - \ell) \theta} e^{i \omega t},
\end{equation}
we find that the coefficients $\alpha_\ell^{(0)}$ and
$\beta_{1-\ell}^{(0)}$ satisfy the eigenvalue equations,
\begin{subequations}
\begin{eqnarray}
\left[ K \left( \ell^2 - \frac{1}{2} \right) + c_1^{\rm 1D} n \right]
\alpha_\ell + c_1^{\rm 1D} n \beta_{1-\ell}^* & = & \hbar \omega
\alpha_\ell,
\nonumber \\
\\
\left[ K \left( \ell^2 + 2 \ell + \frac{1}{2} \right) + c_1^{\rm 1D} n
\right] \beta_{1-\ell}^* + c_1^{\rm 1D} n \alpha_\ell & = & -\hbar
\omega \beta_{1-\ell}^*.
\nonumber \\
\end{eqnarray}
\end{subequations}
Diagonalizing these equations, we obtain the Bogoliubov eigenenergy for
the $m = 0$ excitation as
\begin{equation} \label{Eb}
\hbar \omega = K \left( \ell - \frac{1}{2} \right) + \sqrt{K \ell (\ell -
1) \left[ K \ell (\ell - 1) + 2 c_1^{\rm 1D} n \right]}.
\end{equation}

For $\ell = 0$ or $1$, Eq.~(\ref{Eb}) is always real and there is no
dynamical instability.
For other values of $\ell$, the square root of Eq.~(\ref{Eb}) is imaginary
when $2 c_1^{\rm 1D} n < -K \ell (\ell - 1)$, and the corresponding mode
is dynamically unstable.
The most unstable modes are $\ell = 2$ and $-1$.
As in Eq.~(\ref{three}), these modes correspond to the threefold domain
formation.
Thus, the transverse magnetization is most unstable against forming the
threefold domains, which agrees with the 2D numerical result in
Fig.~\ref{f:dynamics}.
Similarly, the $\ell = 3$ and $-2$ modes correspond to fivefold domains, 
the $\ell = 4$ and $-3$ modes correspond to sevenfold domains, and so on.
In general, dynamical instabilities forming $j$-fold domains with an odd
integer $j \geq 3$ can exist.

Performing the Bogoliubov analysis for $\delta \psi_{\pm 1}$ in a similar
manner, we obtain the eigenenergies as
\begin{widetext}
\begin{eqnarray} \label{Eb1}
\hbar \omega & = & K (\ell - 1) 
\nonumber \\
& & + \biggl\{ K (\ell - 1)^2 \biggl[ K (\ell^2
- 2 \ell + 2) + (c_0^{\rm 1D} + c_1^{\rm 1D}) n
\pm 2 \sqrt{K \left[ K (\ell -
1)^2 + (c_0^{\rm 1D} + c_1^{\rm 1D}) n \right] + (c_0^{\rm 1D} - c_1^{\rm
1D})^2 n^2 / 4} \biggr] \biggr\}^{1/2},
\nonumber \\
\end{eqnarray}
\end{widetext}
where the corresponding mode has a form similar to Eq.~(\ref{mode1}) with
respect to $\ell$.
For $\ell = 1$, there is no dynamical instability, since $\delta \psi_{\pm
1}$ have the same angular momenta as $\psi_{\pm 1}^{\rm hqv}$ in
Eq.~(\ref{1Dhqv}).
For $\ell = 0$ and $2$, Eq.~(\ref{Eb1}) always has an imaginary part for
$c_1^{\rm 1D} < 0$, in agreement with the 2D result in Fig.~\ref{f:Bogo}.
The imaginary part is expanded as $(c_0^{\rm 1D} c_1^{\rm 1D})^{1/2} n +
O(n^2)$.

\section{Geometrical meaning of the threefold domains}
\label{s:geometry}

Now we consider the physical interpretation of the threefold domain
formation.

\begin{figure}[t]
\includegraphics[width=9cm]{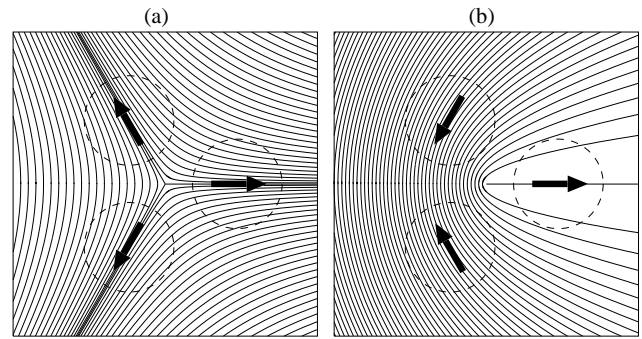}
\caption{
Lines indicating the directions of maximum fluctuations in the transverse
magnetization for the half-quantum vortex state in Eq.~(\ref{af}), where
(a) and (b) correspond to the $+$ and $-$ signs, respectively.
The dashed circles show schematics of the threefold domains and the arrows
show the directions of magnetization in the domains.
}
\label{f:fluc}
\end{figure}

A spin state
\begin{equation} \label{af}
(\zeta_1, \zeta_0, \zeta_{-1}) = (e^{\pm i \theta} \sin \chi, 0, \cos
\chi)
\end{equation}
has spin fluctuations as
\begin{eqnarray}
\Delta f_z^2 & = & \sum_{m,m'} \zeta_m^* (f_z^2)_{mm'} \zeta_{m'} - 
\left[ \sum_{m,m'} \zeta_m^* (f_z)_{mm'} \zeta_{m'} \right]^2 
\nonumber \\
& = & \sin^2 2 \chi, \\
\Delta f_\phi^2 & = & \frac{1}{2} \left[ 1 + \cos(\pm \theta + 2 \phi)
\sin 2\chi \right],
\end{eqnarray}
where $\theta$ is the azimuthal angle ${\rm arg}(x + i y)$ and $f_\phi =
f_x \cos\phi + f_y \sin\phi$.
The transverse fluctuation $\Delta f_\phi^2$ then becomes maximum for
$\phi = \mp \theta / 2$ and $\phi = \mp \theta / 2 + \pi$.
The spatial distributions of the transverse fluctuation exhibit patterns,
as shown in Fig.~\ref{f:fluc}, where the direction of the line indicates
the direction of the maximum transverse fluctuation.

Spontaneous magnetization tends to occur in the directions of large spin
fluctuations, i.e., the directions of the lines in Fig.~\ref{f:fluc}.
For spontaneous magnetization, the total magnetization $\int d\bm{r}
\bm{F}$ must be conserved.
From these two constraints, we understand the reason for the threefold
domain formation.
The arrows in Fig.~\ref{f:fluc} show examples of transverse magnetization
satisfying the two constraints, where the magnetization in each domain
occurs in the direction of the line and the sum of the three magnetization
vectors vanishes.

We note that continuous magnetization for all $\theta$, as in the
polar-core and Mermin-Ho vortices, is impossible, since the symmetry group
of the ferromagnetic state is different from that of the spin state in
Eq.~(\ref{af}).
The half-quantum vortex structure is peculiar to the latter symmetry
group.
Twofold domain formation is also impossible, since spin directions at
$\theta$ and $\theta + \pi$ differ by $\pi / 2$ (not $\pi$) and the total
spin is not conserved.

For $j$-fold domain formation, the center of each domain is located at
$\theta = 2 \pi p / j$ with $p = 0, 1, \cdots, j - 1$.
For each domain, there are two possible directions of magnetization,
$\mp \pi p / j$ and $\mp \pi p / j + \pi$, where the $\mp$ signs
correspond to the $\pm$ signs in Eq.~(\ref{af}).
For the spin conservation to be satisfied, the sum of these magnetization
vectors must vanish:
\begin{equation} \label{cond}
\sum_{p = 0}^{j - 1} \left( e^{\mp i \pi p / j} \;\;\; {\rm or} \;\;\;
e^{\mp i (\pi p / j + \pi)} \right) = 0.
\end{equation}
For $j = 1$ and $2$, Eq.~(\ref{cond}) cannot be satisfied.
For $j = 3$, we find $1 + e^{\mp i (\pi / 3 + \pi)} + e^{i 2 \pi / 3} =
0$, which corresponds to the arrows in Fig.~\ref{f:fluc}.
In general, Eq.~(\ref{cond}) can only be satisfied for an odd $j \geq 3$.
This result agrees with that in Sec.~\ref{s:ring}.

\section{Conclusions}
\label{s:conclusion}

We have studied the dynamics of spontaneous magnetization of a
half-quantum vortex state in a spin-1 BEC with a ferromagnetic
interaction.
Solving the GP equation numerically, we found that the axisymmetry is
spontaneously broken and the threefold magnetic domains are formed through
the dynamical instability (Fig.~\ref{f:dynamics}).
The critical strength of the interaction for the dynamical instability was
obtained by the Bogoliubov analysis (Fig.~\ref{f:Bogo}).
In order to understand the phenomenon in an analytic manner, we
investigated the 1D ring model and showed that the transverse
magnetization is most unstable against forming the threefold domains among
the $j$-fold domains with odd integers $j \geq 3$ (Sec.~\ref{s:ring}).
We provided a physical interpretation of the phenomenon based on the
topological spin structure of the half-quantum vortex and spin
conservation (Fig.~\ref{f:fluc}).

The half-quantum vortex in a spin-1 BEC is peculiar to the symmetry group
that the spin state (\ref{af}) possesses.
For the ferromagnetic interaction, in which the state (\ref{af}) is
unstable, the system exhibits nontrivial dynamics, namely, threefold
domain formation.
We expect that various pattern formation phenomena may occur in magnetic
phase transitions in spinor BECs containing topological structures, in
which the symmetry groups of the spin states change in the phase
transitions.

\begin{acknowledgments}  
This work was supported by the Ministry of Education, Culture, Sports,
Science and Technology of Japan (Grants-in-Aid for Scientific Research,
No.\ 17071005 and No.\ 20540388) and by the Matsuo Foundation.
\end{acknowledgments}

\end{document}